\documentclass[graybox]{svmult}
\usepackage{mathptmx} 
\usepackage{helvet} 
\usepackage{courier} 
\usepackage{type1cm} 
\usepackage{makeidx} 
\usepackage{graphicx} 
\usepackage{multicol} 
\usepackage[bottom]{footmisc}
\makeindex 
\begin{document}

\title*{Modeling of high Reynolds number flows with solid body rotation or magnetic fields}
 \titlerunning{Modeling of anisotropic turbulent flows with either magnetic fields or imposed rotation} 
\author{A. Pouquet, J. Baerenzung, J. Pietarila Graham, P. Mininni, H. Politano \& Y. Ponty}

\authorrunning{A. Pouquet, J. Baerenzung, J. Pietarila Graham, P. Mininni, H. Politano and Y. Ponty}
\author{A. Pouquet\inst{1} \and J. Baerenzung \inst{1} \and J. Pietarila Graham\inst{2} \and P. Mininni\inst{3} \and H. Politano\inst{4} \and Y. Ponty\inst{4}}


\institute{National Center for Atmospheric Research, PO Box 3000, Boulder CO-80307, USA:\texttt{\{pouquet,baeren\}@ucar.edu}
\and  
MPI  f\"ur Sonnensystemforschung, 37191 Katlenburg, Germany:\texttt{jpietarilagraham@mailaps.org}
\and
Departamento de F\'\i sica, Facultad de Ciencias Exactas y Naturales, Universidad de Buenos Aires, Ciudad Universitaria, 1428 Buenos Aires, Argentina, and NCAR:\texttt{mininni@ucar.edu}
\and
Observatoire de la C\^ote D'Azur, Nice, France:\texttt{\{politano,ponty\}@oca.eu}
}


\maketitle

\abstract{We present two models for turbulent flows with periodic boundary conditions and
with either rotation, or a magnetic field in the magnetohydrodynamics (MHD) limit. One model, based on Lagrangian averaging, can be viewed as an invariant-preserving filter, whereas the other model, based on spectral closures, generalizes the concepts of eddy viscosity and eddy noise. These models, when used separately or in conjunction, may lead to substantial savings for modeling high Reynolds number flows when checked against high resolution direct numerical simulations (DNS), the examples given here being run on grids of up to $1536^3$ points.}
 
\abstract{We present two models for turbulent flows with periodic boundary conditions and
with either rotation, or a magnetic field in the magnetohydrodynamics (MHD) limit. One model, based on Lagrangian averaging, can be viewed as an invariant-preserving filter, whereas the other model, based on spectral closures, generalizes the concepts of eddy viscosity and eddy noise. These models, when used separately or in conjunction, may lead to substantial savings for modeling high Reynolds number flows when checked against high resolution direct numerical simulations (DNS), the examples given here being run on grids of up to $1536^3$ points.}

\section{The Lagrangian model } 

Turbulence modeling, in engineering as well as for geo- and astrophysics, is a needed approach even though the power of computers is ever increasing, simply because the number of excited modes in such flows vastly exceeds the capacity of computers in the foreseeable future. As the Reynolds number of DNS grows, tests can be devised which study in detail the properties of such models and thus allow improvements, or else generalizations to handle more complex flows, for example taking into account anisotropies in the presence of either rotation or magnetic fields.

\begin{figure}[htbp]\begin{center}\leavevmode \centerline{%
 \begin{tabular}{c@{\hspace{-.52in}}cc}
    \includegraphics[width=3.4cm]{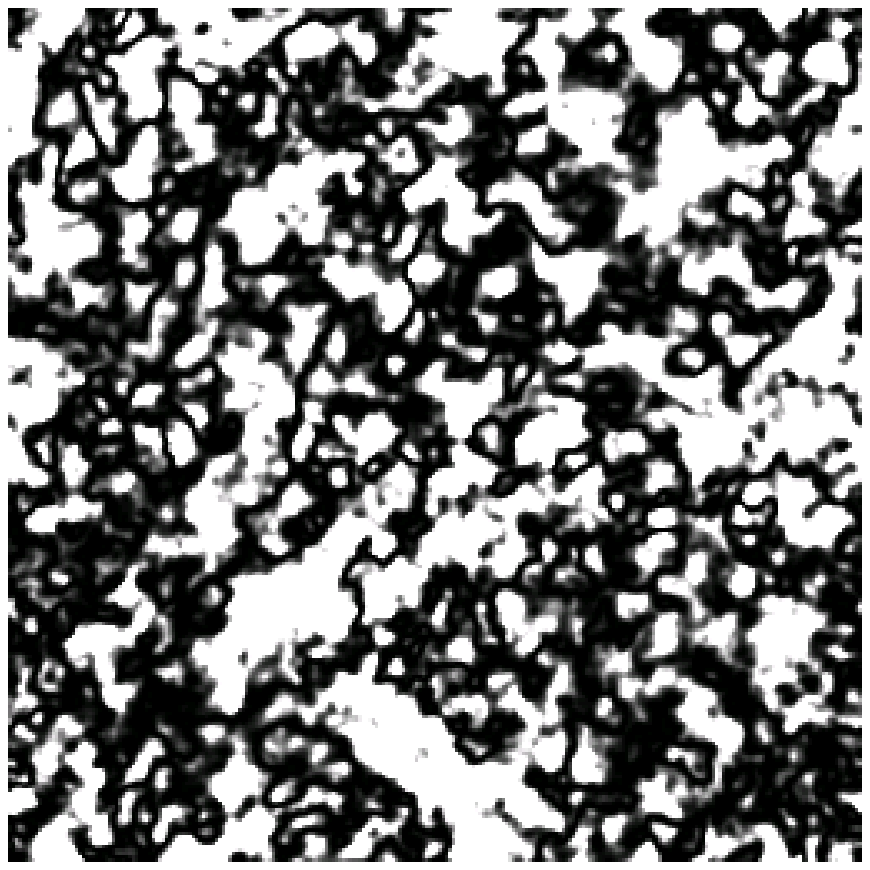} &
 \hskip0.55truein \includegraphics[width=3.6cm]{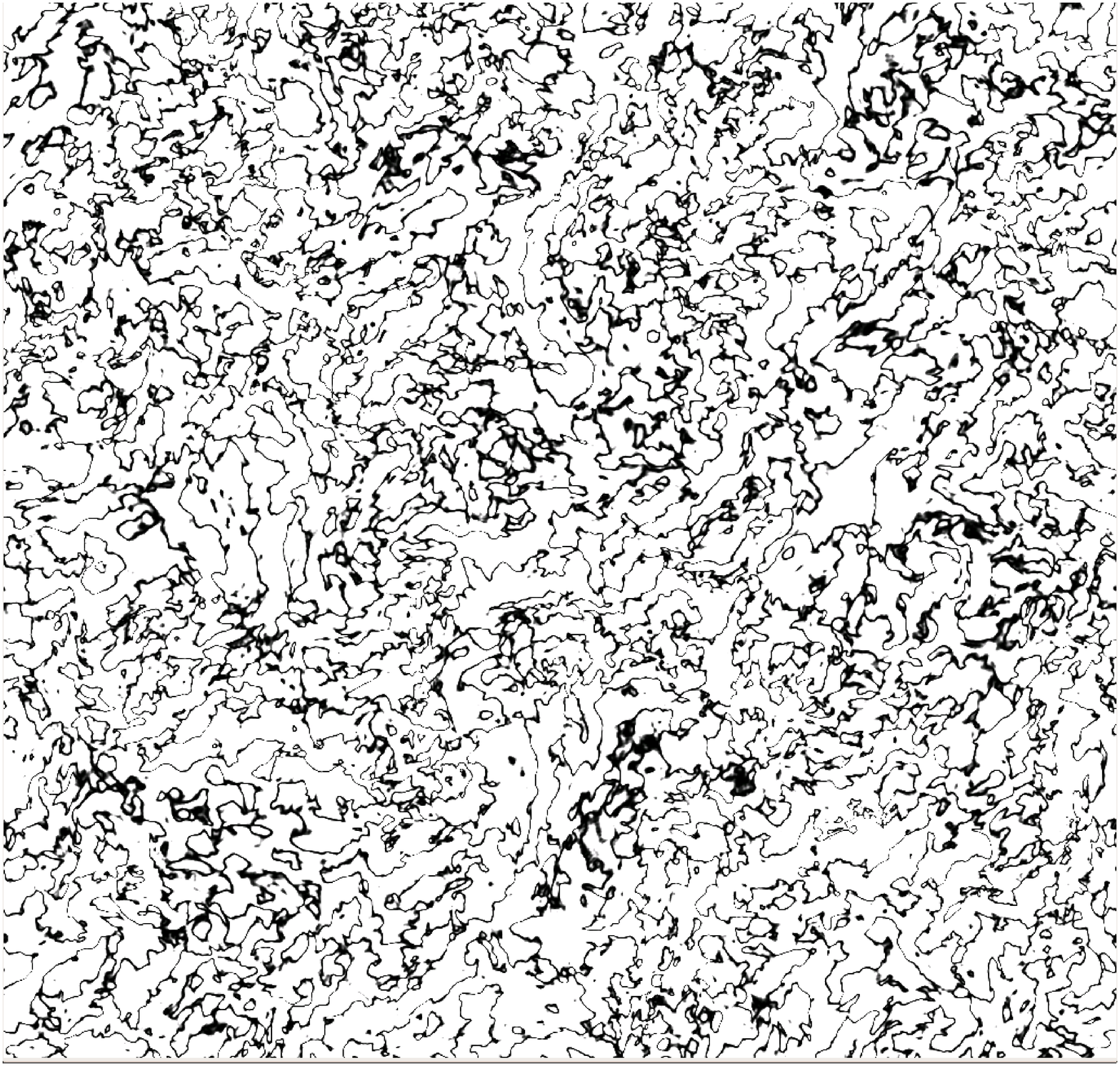} &
  \includegraphics[width=4.9cm]{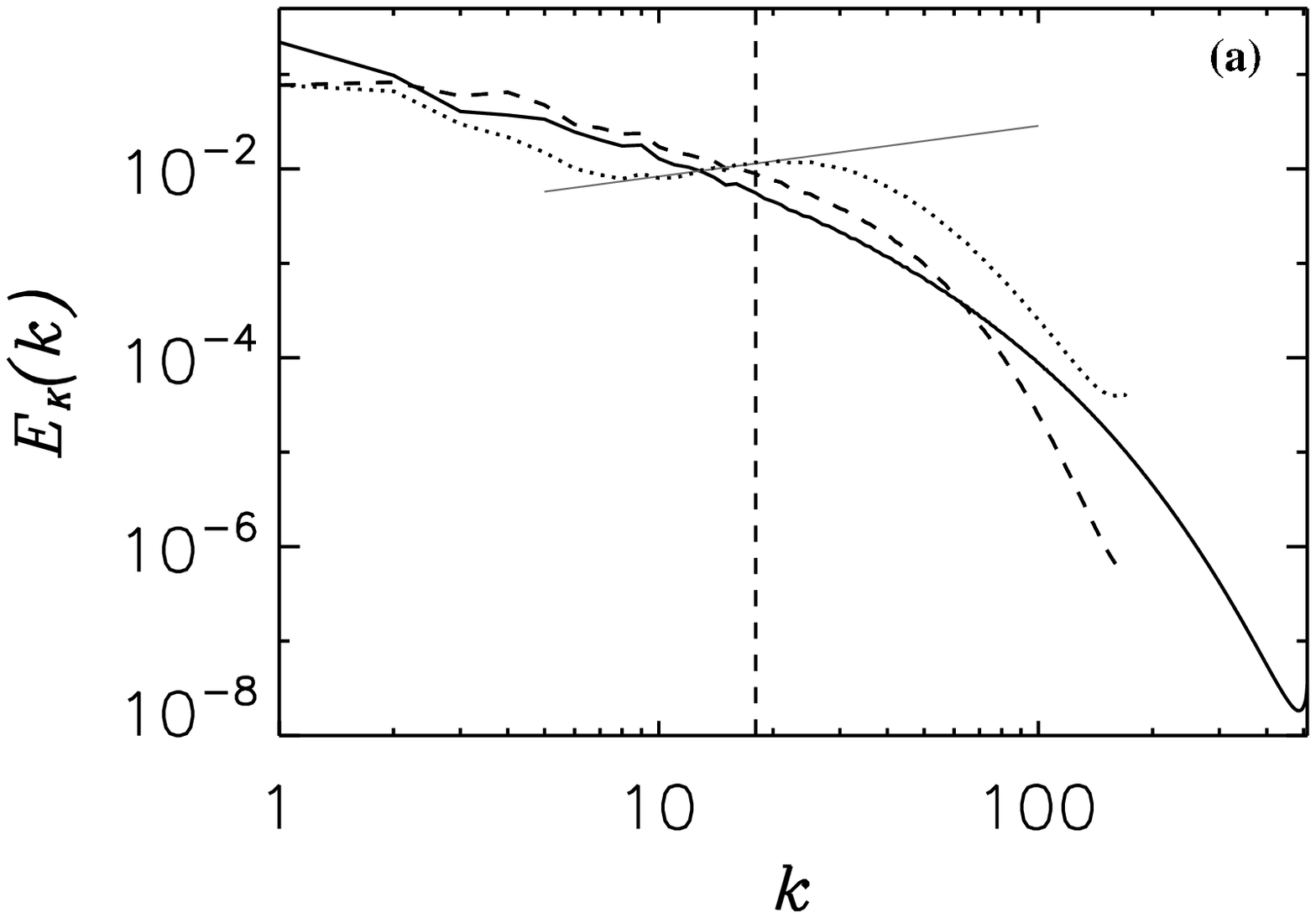} \end{tabular}}
  \caption{
A slice of a region of space in which regions of energy transfer smaller than $1\%$ its mean are shown in black for {\it (left)} a LAM model for fluids on a grid of $256^3$ points and {\it (middle)} a DNS of the Navier-Stokes equations on a $1024^3$ grid at the same Reynolds number: LAM has a substantial reduction in energy transfer, and thus of dissipation, leading to an energy accumulation at small scales, as shown in its energy spectrum.
{\it Right:} kinetic energy spectra for a
$1536^3$ DNS of MHD (solid line), a $512^3$ LAMHD (dash) with filter at $k_\alpha=18$ (vertical dash line), and a $512^3$ LAM (dots), in the latter case with no magnetic field
(${\bf b}\equiv 0$ at all times) 
but otherwise identical LES run. 
For $k \in [5,40]$, LAMHD reproduces well the
scaling of the DNS, with no bottleneck. 
For $k$ close to the filter ($k \in [k_\alpha/2,k_\alpha]$), a $k^{0.5}$ power law (gray line) obtains for fluids using LAM, corresponding to the energy accumulation at small scale for lack of dissipation, whereas it is not present for LAMHD; the magnetic energy has no accumulation of energy at small scale either \cite{jonat_mhd}.
}  \label{FIG:COMP1} \end{center} \end{figure}

\begin{figure}[htbp]\begin{center}\leavevmode \centerline{%
  \begin{tabular}{c@{\hspace{.08in}}c}
  \includegraphics[width=5.9cm]{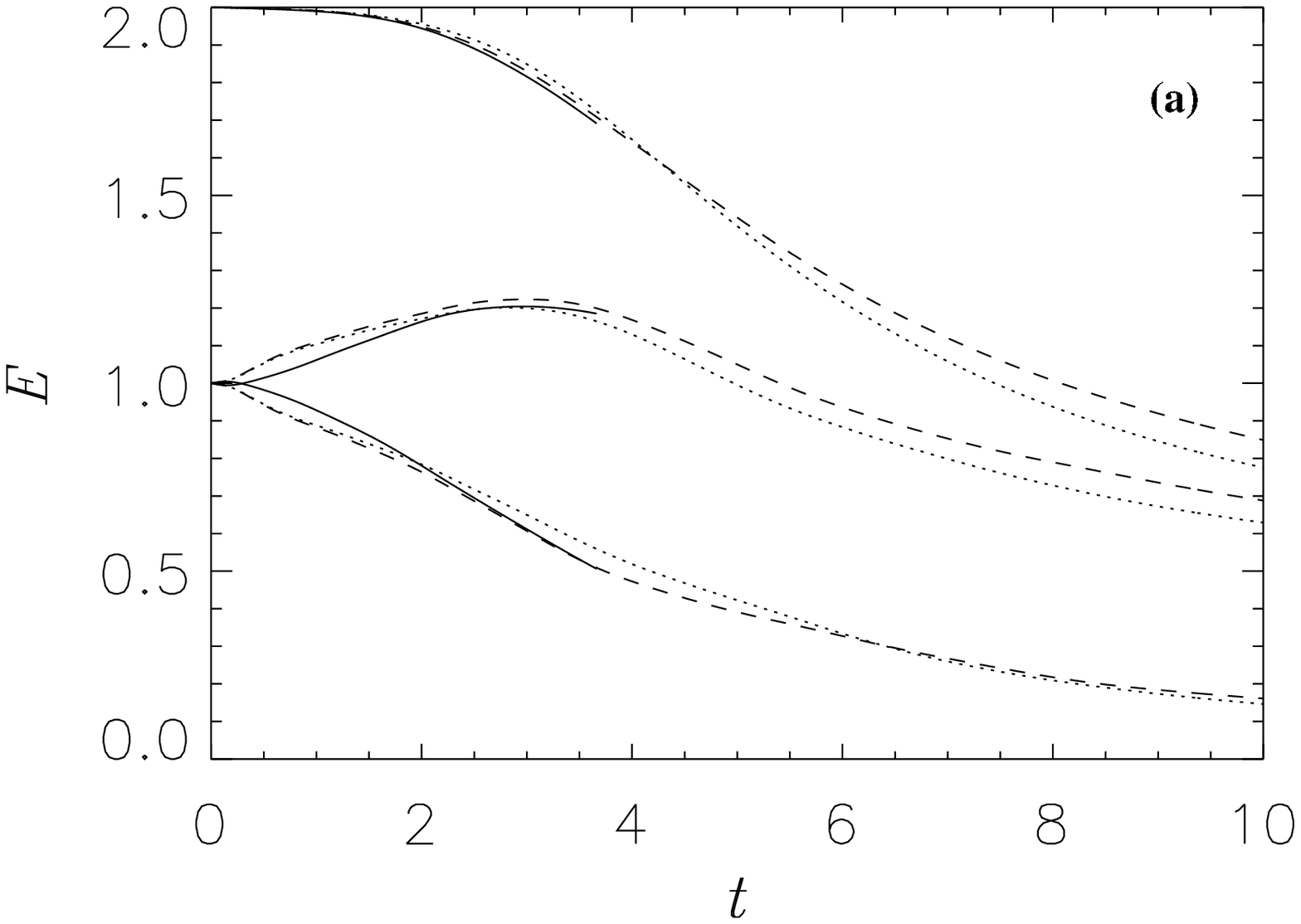} &
  \includegraphics[width=5.9cm]{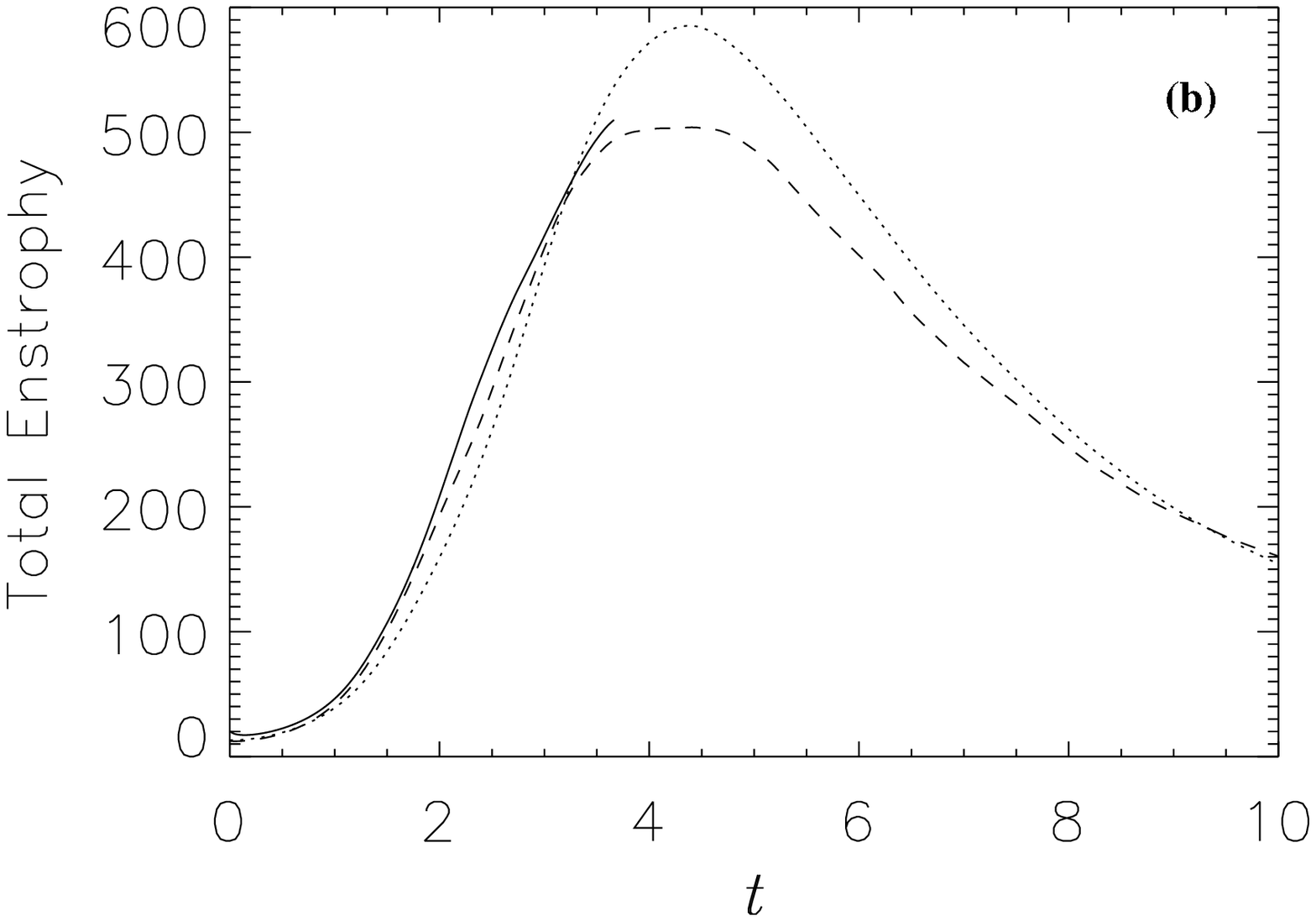}
   \end{tabular}}
  \caption{Temporal evolution of energy (left) and $<J^2+\omega^2>$ (right), where ${\bf \omega}=\nabla \times {\bf v}$
  and ${\bf J}=\nabla \times {\bf b}$, with ${\bf v}$ and ${\bf b}$ the velocity and magnetic field; total energy on top, kinetic $E_V$ and magnetic $E_M$ energies below, with $E_M \ge E_V \ \forall t$.
The thick solid line and dots are DNS on grids of $1536^3$ and $256^3$ points respectively, and the dash line is LAMHD on $256^3$ points, all with the same Reynolds numbers. Only the lower resolution computations are performed beyond the peak of dissipation. 
}
  \label{FIG:COMP2} \end{center} \end{figure}

The first model we have considered can be constructed as a particular filter of small scales \cite{montgo} that preserves invariants of the ideal case but in a different norm ($H_1$ instead of $L_2$). It is called the alpha or Lagrangian averaged model (LAM) \cite{holm}-\cite{holm4} and has been tested in a variety of conditions both in two and three dimensions \cite{chen}--\cite{jonat_mhd2} for Navier-Stokes and MHD. This model can be viewed as a quasi-DNS insofar as it does not introduce by hand a model of the physical effects of the small scales that are neglected, but rather it preserves the Hamiltonian structure of the underlying equations. However, when leaving sufficient room between the filter length $\alpha$ of the model and the smallest resolved scale in the computations, a peculiar feature is observed, namely that small scales are insufficiently dissipated due to a tendency of the model to create regions in space where the normalized energy transfer $\epsilon$ is 
 negligible. This is shown in Fig. \ref{FIG:COMP1} mapping
$\epsilon$ when below 1\% its mean (respective filling factors of regions with negligible $\epsilon$ are $0.26$ and $0.67$ for DNS and LES-LAM).
Note the larger and more numerous patches of negligible transfer in LAM (left) compared to the DNS (middle), leading to an energy accumulation at small scale in the energy spectrum (right, dotted line) before the $\alpha$ cut-off with a positive slope corresponding to a ``bottleneck'' \cite{jonat_NS}.
This bottleneck is absent in MHD: we observe in Fig. \ref{FIG:COMP1} (right) the agreement between the spectra for the MHD-DNS and LAMHD both above and below the filter scale
($\alpha=2\pi/18$); this is probably due to the nonlocality of nonlinear interactions in MHD, and this lack of accumulation of energy observed in the energy spectrum at large wavenumbers represents a marked improvement for the Lagrangian model in MHD when compared to the Navier-Stokes case
\cite{jonat_mhd2}.
 Thus, LAMHD is able to reproduce a DNS on a grid of $1536^3$ points, with savings in CPU and memory usage by a factor of $6$ in linear resolution. In fact, one can pursue the DNS run with LAMHD up to times unreachable with reasonable resources using a DNS (it would take $\sim 1.7 \times 10^6$ CPU hours with present day computers) \cite{jonat_mhd}. In so doing, we observe in Fig. \ref{FIG:COMP2} that equipartition between kinetic and magnetic energy, imposed at $t=0$, is broken in time, with the latter being enhanced by nonlinear interactions; note that the DNS has a small ideal phase where energy is almost conserved and no kinetic-magnetic exchanges take place globally, whereas both the LAMHD and the under-resolved DNS depart from equipartition almost immediately. When considering the total enstrophy (right), the under-resolved run overestimates it because of an accumulation of small-scale excitation not being properly dissipated, whereas LAMHD is much closer to the DNS dat
 a (with a slight under-estimation of it near the peak). 
LAMHD should thus prove quite useful, since it is also known to reproduce well the generation of magnetic fields by velocity gradients (dynamo effect) and the inverse cascade of magnetic helicity, 
as well as small-scale properties such as the variation of the cancellation exponent of the current density in two space dimensions \cite{jonat_cancel}.

\section{Spectral models for rotating flows}

\begin{figure}[htbp]\begin{center}\leavevmode \centerline{%
  \begin{tabular}{c@{\hspace{.15in}}c}
       \includegraphics[width=5.5cm]{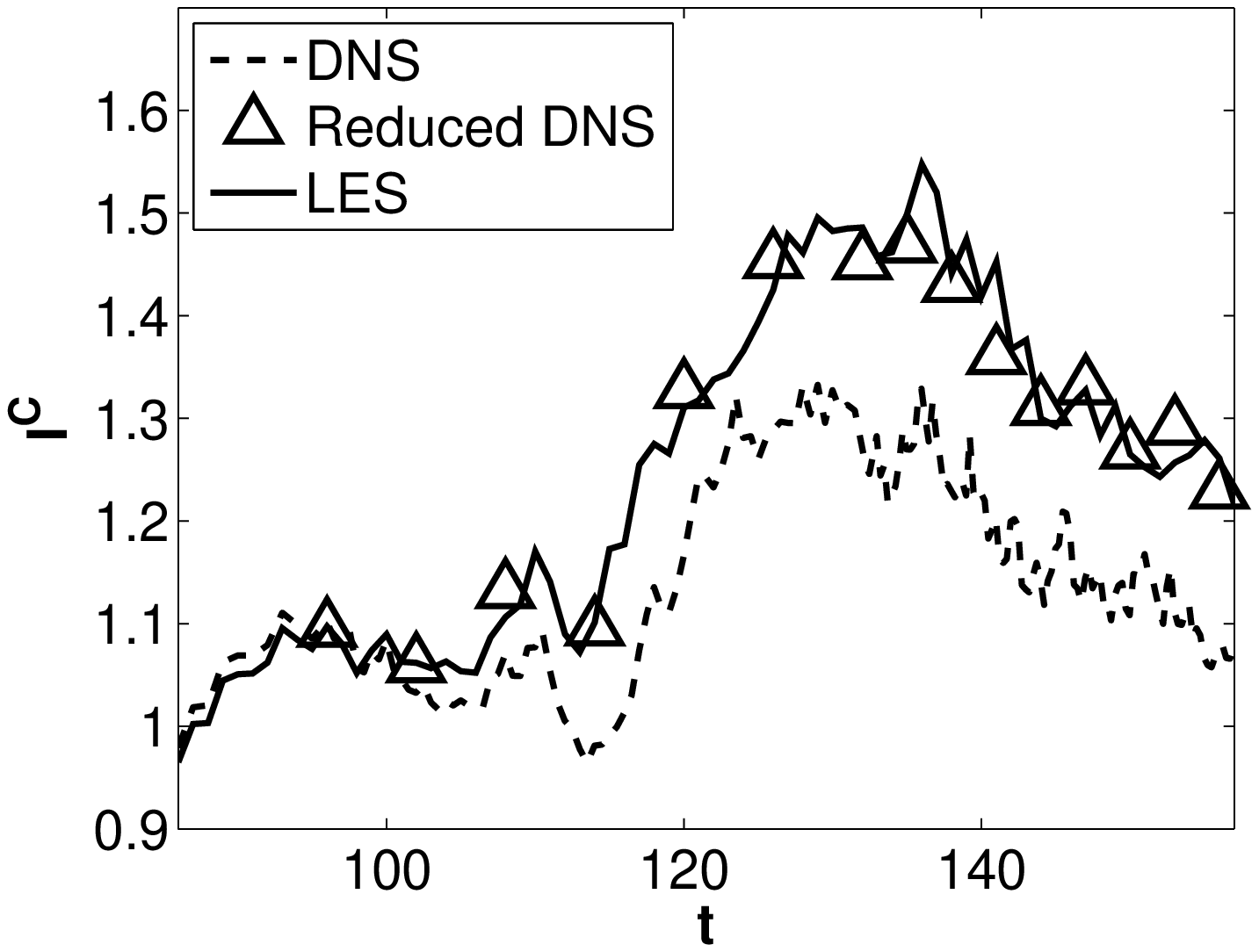} &
   \includegraphics[width=6.55cm]{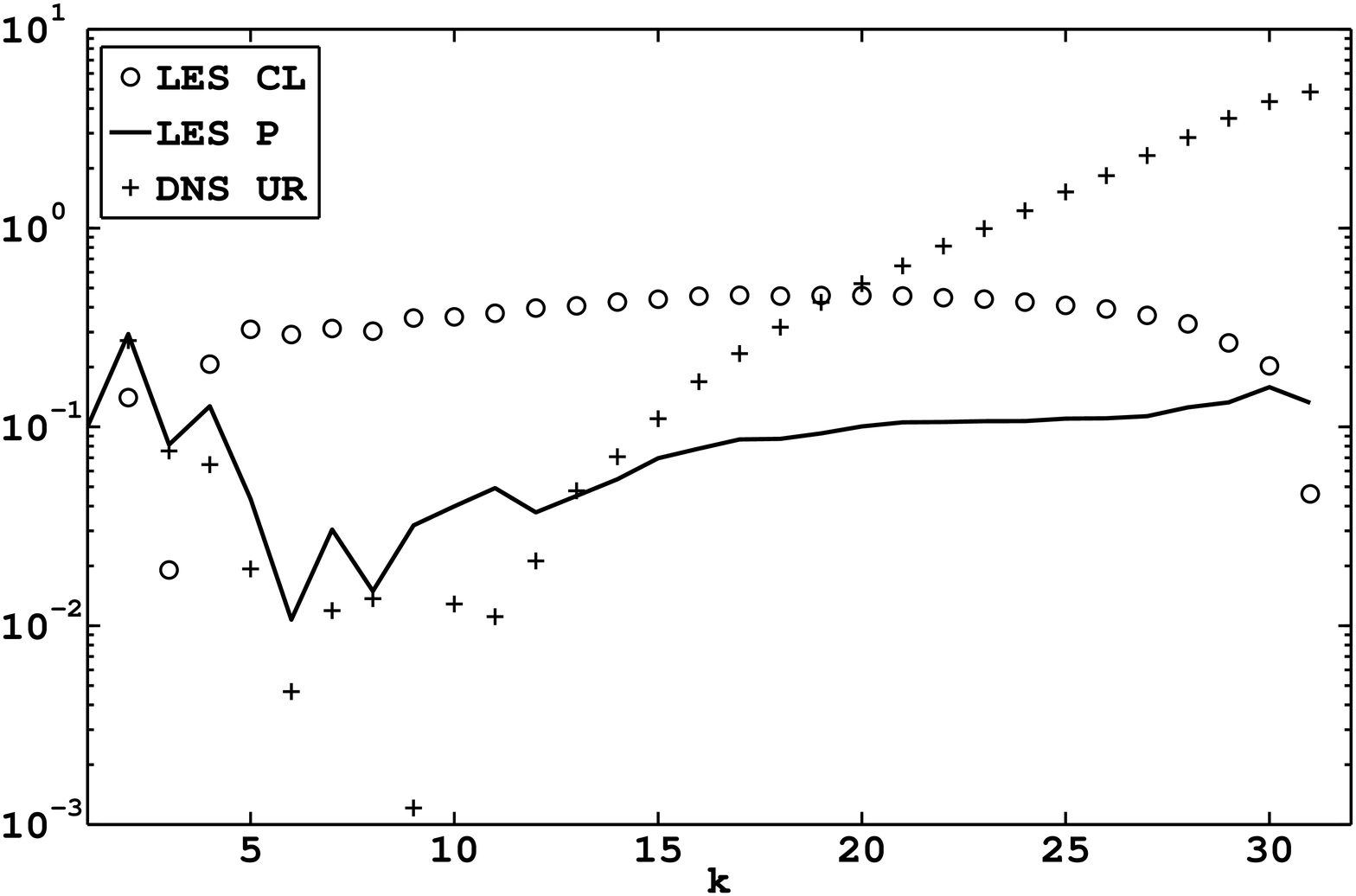} 
 \end{tabular}}
  \caption{
  {\it Left:} time variation (in units of the eddy turn-over time) of the isotropy coefficient
$I^C=\sqrt{<|{\bf v}_{1}|^2>/<|{\bf v}_{2}|^2>}$, with ${\bf v}_{1,\ 2}$ the velocity projected onto ${\bf e}_1= {\bf k}\times {\bf z}$ and ${\bf e}_2={\bf k}\times {\bf e}_1$,
${\bf k}$ being the wavenumber and ${\bf z}$ the axis of rotation;
solid line: LES ($64^3$ grid points); dash: full DNS ($256^3$ points); triangle: full DNS data downgraded to $64^3$ points; Rossby number of $0.03$ with a non-helical forcing at large scale and a non-helical spectral model \cite{julien_rot1}. Note the DNS/LES agreement and the progressive return to isotropy.
{\it Right:} Energy spectrum of the error $[E_{model}-E_{DNS}]/E_{DNS}$ when comparing the DNS on a grid of $1536^3$ points for a rotating flow with $Ro\sim 0.03$ with either an under-resolved DNS (+) on a $160^3$ grid, with the Chollet-Lesieur model \cite{CL} (circles) and the LES we propose here \cite{julien_NS+MHD, julien_rot1} (solid line), both on grids of $96^3$ points. 
The error, exponential at high k for the under-resolved run, is the lowest almost consistently for the spectral model.
} \label{FIG:COMP3} \end{center} \end{figure}

The second model we test in this paper is based on a two-point closure of turbulence, the Eddy Damped Quasi Normal Markovian or EDQNM (see, e.g., \cite{orszag}). In this approach, eddy viscosity and eddy noise 
are included, and the model allows for taking into account non-Kolmogorovian energy spectra to assess these transport coefficients \cite{julien_NS+MHD}--\cite{julien_rot2}; the model builds on the so-called Chollet-Lesieur formulation of spectral eddy viscosity \cite{CL} (hereafter, CL) which is also tested separately against DNS at the same Reynolds number $R_v$ and down to Rossby numbers of $Ro\sim 0.03$. Moreover, helical contributions to the transport coefficients, 
following the helical EDQNM developed in \cite{andre_lesieur}, can be incorporated in the model; these contributions depend on the helicity spectrum at small scale (where the helicity $H_V$ is defined as usual as $\left<{\bf v} \cdot {\bf \omega}\right>$ with ${\bf \omega}=\nabla \times {\bf v}$ the vorticity). For example, $H_V(k)$ being the helicity spectral density, one can write, in the temporal variation of the energy spectrum, the small-scale contributions as:

\begin{equation}
\partial_t E(k)\sim -2k^2E(k) [ \nu + \nu_{turb} ] \  -2 k^2 H(k) \tilde \nu_{turb} \ ;
\label{nu_turb}\end{equation}
Eq. (1) uses a short-hand but hopefully self-explanatory notation to bring the structure of the model (see \cite{julien_NS+MHD} for details), and it omits both the resolved scale contributions and the eddy-noise contributions for simplicity. 
The classical EDQNM eddy viscosity $\nu_t (k,t) \sim \int_{>} f_1(k,p,q)E(q)dpdq$ depends on an integral of the energy spectrum in the small scales 
(symbolized by $\int_{>}$) and represents the drain of energy due to the unresolved sub-grid scales; similarly, $ \tilde \nu_t (k,t) \sim \int_{>} f_2(k,p,q)H_V(q)dpdq$ gives the contribution of small-scale helicity (with ${\bf k}={\bf p}+{\bf q}$ due to the convolution).

 making it more difficult for the dynamo effect to take place and thus leading to a higher value of the critical parameter $R_M^C$.Note that overlap between the different methods also allows for an inter-assessment of the models. What is still not entirely clear is whether, for low $P_M$, a plateau obtains (as the new data coming from the spectral model seems to indicate) or whether lower values of $R_M^C$ are to be expected, as one approaches realistic values of $P_M$ ($\sim 10^{-6}$) and as found using the EDQNM model by itself \cite{leorat} (as opposed to incorporating it in a LES, as reported here).
 
When modeling rotating flows (here, with no magnetic fields) \cite{cambon_book}, one introduces an obvious external anisotropy in the problem. However, anisotropic models, as those developed using for example extensions of EDQNM to such flows (see e.g. \cite{cambon_scott}-\cite{cui} 
 for recent works) are costly since they now depend on both the parallel and perpendicular (referring to the rotation axis) components of the velocity. On the other hand, one can remark that in an LES approach, one models the small scales which can recover some degree of isotropy since the scale-dependent Rossby number $Ro=v_{\ell}/\ell \Omega$ with $v_{\ell}$ the velocity at scale $\ell$, and $\Omega$ the rotation rate, gets larger as $\ell \rightarrow 0$.
  In Fig. \ref{FIG:COMP3} (right) is given the temporal evolution of the isotropy coefficient $I^C$ (see caption for definition) for the full DNS for a run forced with a non-helical velocity field (the Taylor-Green flow), the DNS data being downgraded to the grid resolution of the LES and the LES using the spectral model we propose \cite{julien_rot2}; this coefficient, of unit value for full isotropy (rotation is introduced in the run at $t\sim 90$ after the flow has settled to a turbulent state), begins to increase substantially once the inverse cascade of energy builds up, for $t \ge 110$, and then decreases under the influence of the small-scale cascade that restores isotropy to some extent.
Furthermore, Fig. \ref{FIG:COMP3} shows that the LES, when compared to the DNS downgraded to the LES run, reproduces this result quite accurately.
This means that, at least at the moderate Rossby number of these computations, down to $Ro\sim 0.03$ and micro-Rossby number $\omega_{rms}/\Omega \approx 1$ (with $\omega_{rms}$ the rms value of the vorticity), an isotropic approach is a workable solution for modeling such flows since the small scales are sufficiently isotropic. Whether such an agreement will persist at lower Rossby numbers is left for future investigations, but since $Ro\approx 0.1$ in the atmosphere, this spectral model may prove useful in this context. Fig. \ref{FIG:COMP3} (right) confirms this result, by plotting the energy spectral error for three models (see caption).

\begin{figure}[htbp]\begin{center}\leavevmode \centerline{%
  \begin{tabular}{c@{\hspace{.15in}}c}
   \includegraphics[width=6.7cm]{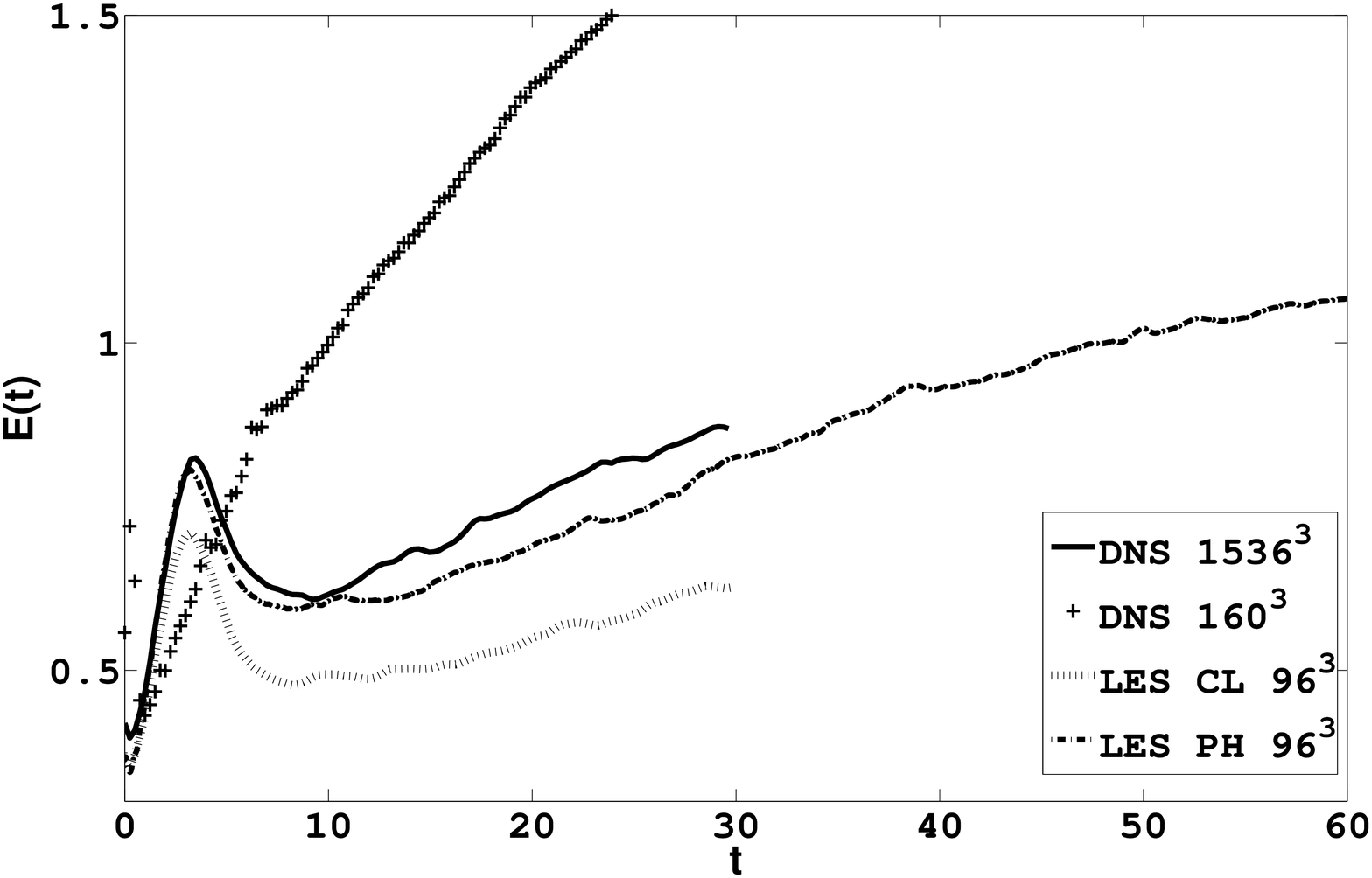} &
     \includegraphics[width=6.7cm]{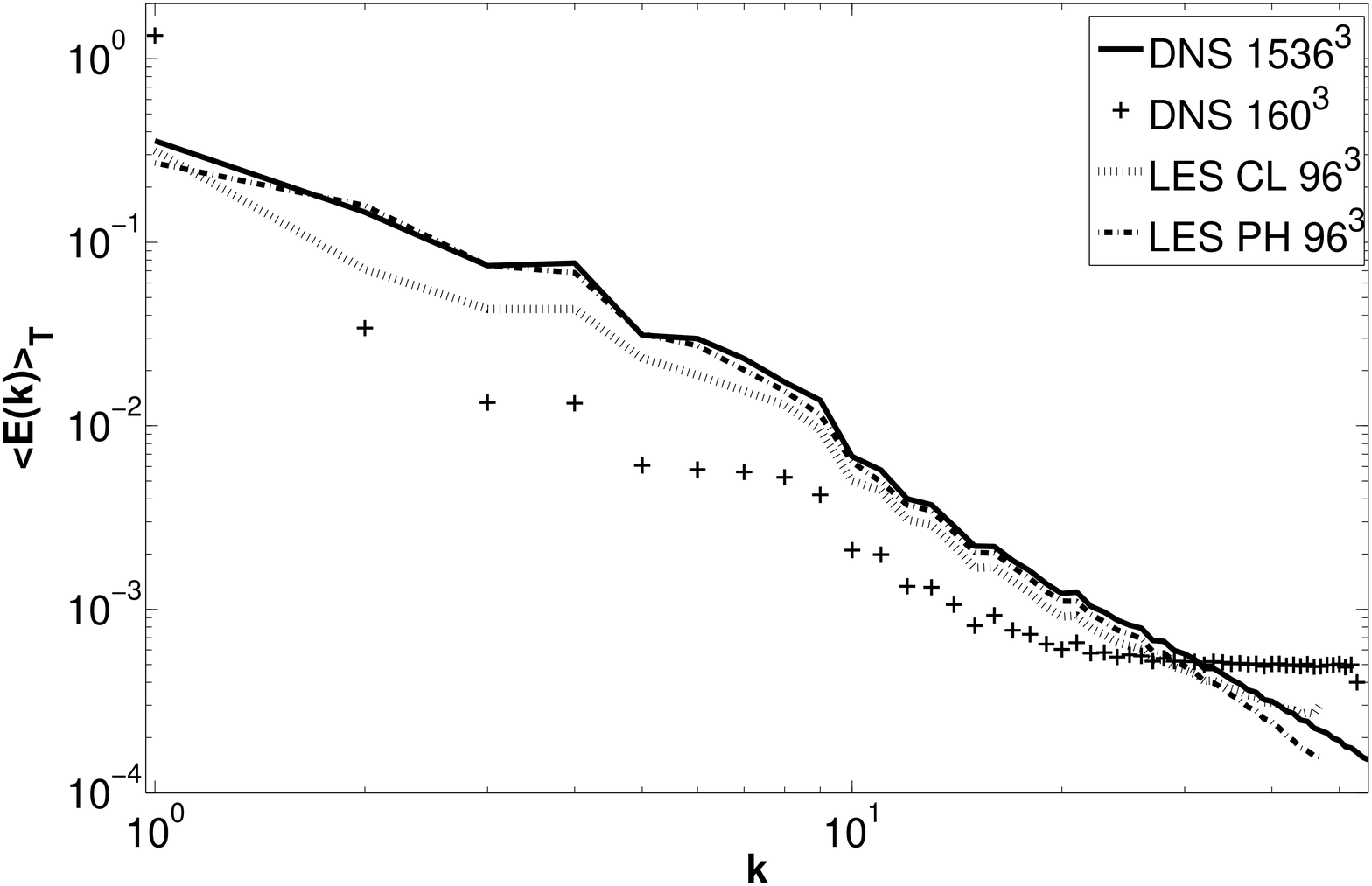}
    \end{tabular}}
  \caption{
Helical rotating flow: comparisons between a DNS on a grid of $1536^3$ points (solid line), an under-resolved $160^3$ DNS (+), the LES-CL model (grey line), and the helical spectral model (LES-PH, dash) \cite{julien_NS+MHD}; the two LES runs use grids of $96^3$ points. The
Reynolds number is 5600 and the Rossby number is 0.06.
{\it Left:} Energy as a function of time; note the unphysical substantial increase in the case of the under-resolved run (+).
{\it Right:} Energy spectrum averaged from $t=20$ to $t=30$; again, the under-resolved run clearly underperforms the LES models, and LES-PH is the model closest to the DNS at a substantial savings in computational cost compared to the DNS.
} \label{FIG:COMP4} \end{center} \end{figure}

Noting that it has been found recently that helicity plays an important role in the dynamics of turbulent flows in the presence of rotation \cite{pablo_rot_hel}, a point that may relate to a simplified dynamics of tornadoes, we test further the possibility of using isotropic spectral models for rotating flows by performing a comparison against a massive DNS of a rotating helical flow, on a grid of $1536^3$ points; note that more than 700,000 CPU hours were used for this second large DNS run; the Beltrami forcing is an ABC flow \cite{ABC} set at wavenumber $k_F=7$, leaving room for both a direct cascade and an inverse cascade to take place. Among the many novel features of such a flow \cite{pablo_rot_1536}, we display here a comparison on two diagnostics, see Fig. \ref{FIG:COMP4}: when examining the temporal evolution of the total energy (left) and the energy spectra averaged over a few turn-over times (right), we see that the LES model (including for this fully helical case,
  the helical contributions to eddy viscosity and eddy noise \cite{julien_NS+MHD}) performs best, and the under-resolved DNS performs worst, in particular because of an accumulation of energy at both small and large scales. The Chollet-Lesieur model obtains a growth rate for the energy in the inverse cascade quite close to that in the DNS  but is somewhat more dissipative, whereas the spectral model behaves better energetically. Similarly, for the energy spectra, the spectral model performs best. 
This data thus provides an unambiguous display of the added value of a LES when contrasted either to under-resolved DNS or to simple eddy-viscosity models in order to approach the dynamics of complex turbulent flows, with here a huge gain in resolution (all LES are performed on a grid of $[1536/16]^3=96^3$ points).

 \section{Conclusion}
 
The increase of power in computers, with the petascale initiative and beyond, does not mean one need not worry about modeling of turbulent flows, quite to the contrary. With increased capability, one will tackle more complex problems with non trivial geometries and micro-physics, as needed in a comprehensive approach to climate, weather, and space physics modeling for example. But because realistic parameters are still well out of range, we can foresee complementary roles for DNS and LES, together with experiments and observations: (i) analysis of the dynamics of complex turbulent flows with highly resolved DNS, followed by (ii) verification and amelioration of models against such DNS runs, the models being used either alone or in a combined fashion (see, e.g., \cite{ponty1} using both LAMHD and LES-PH for the dynamo problem at low magnetic Prandtl number); then (iii) exploration of parameter space with such models, and finally (iv) starting again the cycle with new DNS runs
 . Such a cyclic approach relies on Moore's law of doubling of processor speed every $\approx 18$ months, leading to a doubling of resolution in a 3D run every $\approx 6$ years.
In the specific cases mentioned in this paper, the savings at given Reynolds (and magnetic Prandtl or Rossby) numbers, are already substantial since a LES run on a grid of $96^3$ points reproduces satisfactorily a DNS run that cost almost $10^4$ times more.
Such models thus should prove useful in exploring parametrically dynamical regimes of geophysical and astrophysical turbulence in the presence of rotation and/or magnetic fields in a variety of conditions such as they arise in nature.


\begin{thebibliography}{99}
%

  \bibitem{montgo}
  Montgomery, D., Pouquet, A.:
  An alternative interpretation for the Holm ``alpha'' model.
 Phys. Fluids, {\bf 14}, 3365--3366 (2002)
  
 \bibitem{holm}
 Chen, S.Y.,  Holm, D.D., Margolin, L.G., Zhang, R.:
 Direct numerical simulations of the Navier-Stokes alpha model.
  Physica D, {\bf 133}, 66-83 (1999)

 \bibitem{holm3}
  Foias, C., Holm, D.D. ,  Titi, E.S.:
  The Navier-Stokes-alpha model of fluid turbulence.
   Physica D {\bf 152-153} 505-519 (2001)

 \bibitem{holm4}
Holm, D.D.: 
Lagrangian averages, averaged Lagrangians, and the mean effects of fluctuations in fluid dynamics.
Chaos, {\bf 12}, 518-530 (2002)
   
   \bibitem{chen}
Chen, S.,  et al.:
A connection between the Camassa-Holm equations and turbulent ßows in channels and pipes.
Phys. Fluids {\bf 11}, 2343-2353 (1999)

\bibitem{nadiga}
  Nadiga, B., Shkoller, S.:
  Enhancement of the inverse-cascade of energy in the 2D 
Lagrangian-averaged Navier-Stokes equations.
Phys. Fluids {\bf 13}, 1528-1531 (2001)

  \bibitem{comp_2d_3d}
Mininni, P., Montgomery, D., Pouquet, A.: 
Numerical solutions of the three-dimensional MHD alpha model.
Phys. Fluids {\bf 17}, 035112 (2005)

  
   \bibitem{jonat_mhd}   Pietarila Graham, J. {\it et al.}:   Inertial Range Scaling, K\'arm\'an Theorem and Intermittency for Forced and Decaying Lagrangian Averaged MHD in 2D.  Phys. Fluids. {\bf 18} 045106 (2006)
     
  \bibitem{jonat_NS}
  Pietarila Graham, J. {\it et al.}: 
 Highly turbulent solutions of LANS-$\alpha$ and their LES potential.
 Phys. Rev. E {\bf 76}, 056310 (2007)
  
   \bibitem{jonat_NS2}
  Pietarila Graham, J. {\it et al.}: 
  Three regularization models of the Navier-Stokes equations.
  Phys. Fluids {\bf 20}, 035107 (2008)
  
      \bibitem{jonat_mhd2}
   Pietarila Graham, J., Mininni, P., Pouquet, A.:
``The Lagrangian-averaged model for MHD turbulence and the absence of bottleneck,''
 submitted to Phys. Rev. E (2008) \& arxiv/0806.2054v1
  
  \bibitem{jonat_cancel}
Graham, J.,  Mininni, P., Pouquet, A.:
Cancellation exponent and multifractal structure in Lagrangian 
averaged magnetohydrodynamics.
Phys. Rev. E {\bf 72}, 045301(R) (2005)

\bibitem{orszag}
Orszag, S., Kruskal, M.: Formulation of the theory of turbulence. Phys. Fluids {\bf 11}, 43-60 (1968)

  \bibitem{julien_NS+MHD}
  Baerenzung, J. {\it et al.}:  
 Spectral Modeling of Turbulent Flows and the Role of Helicity. 
  Phys. Rev. E {\bf 77}, 046303 (2008)
  
   \bibitem{julien_NS+MHD2}    Baerenzung, J. {\it et al.}:  ``Spectral Modeling of Magnetohydrodynamic Turbulent Flows,'' 
   Phys. Rev. E {\bf 78}, 026310 (2008)
 

  
  \bibitem{pablo_rot_TG}
Mininni, P., Alexakis, A., Pouquet, A.,
``Scale interactions and scaling laws in rotating flows at moderate Rossby numbers and large Reynolds numbers,'' 
 Phys. Fluids {\bf 21}, 015108 (2009)

\bibitem{pablo_rot_hel}
 Mininni, P., Pouquet, A., ``Helicity cascades in rotating turbulence,''  to appear,
 Phys. Fluids (2009).
See also arxiv:0809.0869

 \bibitem{julien_rot1}  Baerenzung, J., {\it et al.}: ``Modeling of rotating flows at moderate Rossby numbers,'' submitted to Phys. Rev. E  (2009). See also arXiv:0812.1821v1
 
 \bibitem{julien_rot2}
  Baerenzung, J., {\it et al.}: ``Modeling of rotating flows with helicity,'', in preparation (2009)
  
 \bibitem{CL}
 Chollet, J.P., Lesieur, M.:  
Parametrization of small-scale three-dimensional isotropic turbulence using spectral closures. 
 J. Atmos. Sci., {\bf 38}, 2747-2757 (1981)

\bibitem{andre_lesieur}
Andr\'e, J.C.,  Lesieur, M.:
Influence of Helicity on the  Evolution of Isotropic Turbulence at High Reynolds Number.
J. Fluid Mech., {\bf 81}, 187-207 (1977)

\bibitem{leorat} L\'eorat, J., Pouquet, A., Frisch, U.: Fully developed MHD turbulence near critical magnetic Reynolds number.  J. Fluid Mech., {\bf 104},419--443 (1981)

\bibitem{cambon_book}
Sagaut, P., Cambon, C.,  {\it Homogeneous Turbulence Dynamics},  Cambridge Univ. Press 
 (2008)
 
\bibitem{cambon_scott}
Cambon, C., Scott, J.F.:
Linear and nonlinear models of anisotropic turbulence.
Ann. Rev. Fluid Mech. {\bf 31}, 1-53 (1999)

\bibitem{cambon_NJP}
Cambon, C., Rubinstein, R., Godeferd, F.S.:
Advances in wave turbulence: rapidly rotating flows.
New J. Phys. {\bf 6}, Number 73, 1-29 (2004)

\bibitem{cui}
Cui, G.X. et al.:
A new subgrid eddy-viscosity model for 
large-eddy simulation of anisotropic turbulence.
 J. Fluid Mech. {\bf 582}, 377-397 (2007)

\bibitem{ABC}
  Childress S., Gilbert, A.:  {\it Stretch, Twist, Fold: The Fast Dynamo}, Springer-Verlag 
  (1995)

 \bibitem{pablo_rot_1536}  Mininni, P., Pouquet, A.,  ``Persistent cyclonic structures in self-similar turbulent flows,''   submitted to Phys. Rev. Lett. \&  
  arXiv:0903.2294 (2009)

  \bibitem{ponty1}  Ponty, Y. {\it et al.}:  Critical magnetic Reynolds number for dynamo action as a function of magnetic Prandtl number. Phys. Rev. Lett. {\bf 94} 164502 (2005) 

\end{thebibliography}
\end{document}